\documentstyle[12pt]{article}

\newcommand{\bz}{\overline{z}}
\newcommand{\be}{\begin{equation}}
\newcommand{\ee}{\end{equation}}
\newcommand{\ba}{\begin{array}}
\newcommand{\ea}{\end{array}}
\newcommand{\bea}{\begin{eqnarray}}
\newcommand{\eea}{\end{eqnarray}}
\newcommand{\bt}{\begin{tabular}}
\newcommand{\et}{\end{tabular}}
\newcommand{\non}{\nonumber}
\newcommand{\lc}{\left[}
\newcommand{\rc}{\right]}
\newcommand{\lk}{\left\{}
\newcommand{\rk}{\right\}}
\newcommand{\lp}{\left(}
\newcommand{\rp}{\right)}
\newcommand{\ro}{\right.}
\newcommand{\lo}{\left.}
\newcommand{\medio}{\frac{1}{2}}
\newcommand{\cuarto}{\frac{1}{4}}
\newcommand{\superfrac}[2]{\bt{c} $#1$ \\ \hline $#2$ \et}
\newcommand{\eid}{\epsilon_{I}}
\newcommand{\eiu}{\epsilon^{I}}

\newcommand{\eju}{\epsilon^{J}}
\newcommand{\eio}{\epsilon_{I_{0}}}

\begin{document}
\begin{titlepage}

\begin{flushright}
SU-ITP-92-24\\ hepth@xxx/yymmnn\\ \today
\end{flushright}

\begin{center}

\baselineskip25pt
{\LARGE {\bf ELECTRIC-MAGNETIC DUALITY AND SUPERSYMMETRY IN STRINGY
BLACK HOLES}}

\vspace{1cm}

{\large {\bf Tom\'as Ort\'{\i}n}%
\footnote{Bitnet address: tomaso@slacvm}\\
{\it Departament of Physics}\\
{\it Stanford University}\\
{\it Stanford CA 94305, U.S.A.}}

\end{center}

\vspace{1.5cm}

\begin{abstract}

We present a generalization of the $U(1)^{2}$ charged dilaton black
holes family whose main feature is that both $U(1)$ fields have electric
and magnetic charges, the axion field still being trivial. We show
the supersymmetry of these solutions in the extreme case, in which the
corresponding generalization of the Bogomolnyi bound is saturated and a
naked singularity is on the verge of being visible to external observers.
Then we study the action of a subset of the $SL(2,R)$ group
of electric-magnetic duality rotations that generates a non-trivial
axion field on those solutions. This group of transformations is an
exact
symmetry of the $N=4$ $d=4$ ungauged supergravity equations of motion.
It has been argued recently that it could be an exact symmetry of the
full
effective string theory. The generalization of the Bogomolnyi bound is
invariant under the full $SL(2,R)$ and the solutions explicitly rotated
are shown to be supersymmetric if the originals are. We conjecture
that any $SL(2,R)$ transformation will preserve supersymmetry.

\end{abstract}

\end{titlepage}

\newpage

\baselineskip15pt

\section*{Introduction}

Nowadays, a lot of effort is being devoted to the study of black
holes in string theory from different perspectives as an attempt to
elucidate the properties of the quantum gravity theory which is embedded
in it. Many of the unusual features of this theory have their origin
in the atypical coupling between the graviton and the dilaton fields
which has become the particular signature of string theory in toy and
low energy models.

Classical solutions of the low energy effective actions of string theory
in four dimensions with a non-trivial dilaton field are very interesting
in this framework \cite{kn:GHS}\cite{kn:HW}. They exhibit the classical
properties of the dilaton interactions and provide the starting
point for the study of semiclassical (quantum) behavior (Hawking
radiation etc.). This and the fact that they may inherit properties
of the full theory (some unbroken supersymmetries \cite{kn:KLOPP}, some
properties of the spectrum \cite{kn:Sendual} etc.) justify their study.

Another
stimulating aspect of these solutions is that they can always be seen
as solutions of the general relativity theory in the presence of exotic
matter. This naturally suggests the embedding in a locally
supersymmetric theory also hinted by string theory. Local supersymmetry
techniques might be major tools in
the study of general relativity. As an example let us recall Witten's
proof of the positive energy theorem in general relativity \cite{kn:W}
using techniques borrowed from supergravity (see \cite{kn:Hull}
and references therein). After Witten's work, spinor techniques
related to supergravity were used to prove other results of the
same kind, inequalities relating the (asymptotically defined) charges
of spacetimes for which some positivity condition of the energy-momentum
tensor holds (see for example \cite{kn:GH} and the review
paper \cite{kn:Hor}). These inequalities are also known in the context
of supersymmetry as Bogomolnyi bounds \cite{kn:WO}\cite{kn:FSZ} and
solutions that saturate them have special supersymmetric and
geometrical properties. It has been noticed many times that ``extreme"
black holes (i.e. on the verge of showing a naked singularity)
admit unbroken supersymmetries and saturate a Bogomolnyi-like bound (see
\cite{kn:KLOPP} and references therein). Therefore, supersymmetry
seems to act as a cosmic censor. Some exceptions are known in
non-asymptotically flat backgrounds
\cite{kn:Rom}, but we still believe that some connection may exist
between cosmic censorship and supersymmetry in some restricted subset
of asymptotically flat spacetimes. This idea is one of the motivations
for looking at the supersymmetry properties of these solutions.

Finally, methods for generating new solutions from those already
known have recently been described in \cite{kn:Senrot},
\cite{kn:STW} and \cite{kn:Sendual}. They take advantage of
the symmetries of the equations of motion. From the viewpoint of string
theory the non-invariance of the effective  action is not an issue
because the only virtue of the action is that its extrema verify those
equations of motion. We will be interested in the later of this methods,
which consists in performing a combination of ``dual rotations" and
constant shifts of the axion field of a given solution. This operation
rotates electric and magnetic charges and, as we will see, dilaton and
axion charges into each other so we can find
new solutions with non-trivial axion field starting from solutions
with a non-trivial dilaton. These transformations where known to
generate the
$SL(2,R)$ group of exact symmetries of the equations of motion of $N=4$,
$d=4$ ungauged supergravity (the $SU(1,1)$ group of reference
\cite{kn:CSF}) whose action coincides with part of the effective action
of the dimensionally reduced action of the heterotic string. In
reference \cite{kn:Sendual}, A. Sen has shown that this is still a
symmetry of the equations of motion when one includes scalar and
vector terms coming from the compactification process and has
conjectured that $SL(2,Z)$ might be a symmetry of the full effective
string theory after the axion shift symmetry is broken by instantons.
The preservation of unbroken supersymmetries of a quite general
family of extreme spherical charged dilaton black holes by a subgroup
of these transformations is one of the main results of this paper.

The
family of charged dilaton black holes that we are going to study here
generalizes the one recently discussed in \cite{kn:KLOPP} (originally
discovered by Gibbons \cite{kn:Gibb} and discussed by Gibbons and Maeda
\cite{kn:GM}). It will be introduced in section 1. The main feature
of these solutions is the presence of magnetic and electric charges
corresponding to each of the $U(1)$ fields, while the axion remains
trivial. In section 2 we will focus on the supersymmetry properties of
the extreme solutions and will show the existence of supercovariantly
constant spinors in those backgrounds and the saturation
of a Bogomolnyi-like bound. In section 3 we will study the effect
of the subgroup of $SL(2,R)$ used in reference \cite{kn:STW} on the
charges and will see that the corresponding Bogomolnyi-like bound is
left invariant. This will be shown in section 4 to mean that backgrounds
obtained by rotating supersymmetric ones, are also supersymmetric.
Since the full $SL(2,R)$ preserves the Bogomolnyi bound, we conjecture
that the same will happen with any of those transformations.
A discussion and further comments are the content of the last
section and some conventions are specified in the appendix.

\section{Doubly charged solutions}

The truncation of the action of the $SU(4)$ version of $N=4$, $d=4$
ungauged supergravity we are going to work with is
(our conventions are the same as in  \cite{kn:KLOPP})

\bea
S_{SU(4)}                                           & = &
\int dx^{4}\sqrt{-g}\lk-R+2(\partial\phi)^{2}
+\medio e^{4\phi}(\partial a)^{2}-\ro                         \non \\
                                                    &   &
\lo-e^{-2\phi}[F^{2}+G^{2}] +ia[F \star F+ G \star G]\rk.
\eea

The equations of motion read

\bea
\nabla_{\mu}(e^{-2\phi}F^{\mu\nu}-ia\star F^{\mu\nu}) & = & 0, \non \\
\nabla_{\mu}\star F^{\mu\nu}                          & = & 0, \non \\
\nabla_{\mu}(e^{-2\phi}G^{\mu\nu}-ia\star G^{\mu\nu}) & = & 0, \non \\
\nabla_{\mu}\star G^{\mu\nu}                          & = & 0, \non \\
\nabla^{2}\phi-\medio e^{4\phi}(\partial a)^{2}
-\medio e^{-2\phi}(F^{2}+G^{2})                       & = & 0, \non \\
\nabla^{2}a+4\partial_{\mu}\phi\partial^{\mu}a
-ie^{-4\phi}(F\star F+G\star G)                       & = & 0, \non \\
R_{\mu\nu}+2\partial_{\mu}\phi\partial_{\nu}\phi
+\partial_{\mu}a\partial_{\nu}a-                      &   &   \non \\
-2 e^{-2\phi}\lk \lc F_{\mu\rho}F_{\nu}^{\rho} -\cuarto g_{\mu\nu}F^{2}
+G_{\mu\rho}G_{\nu}^{\rho} -\cuarto g_{\mu\nu}G^{2}\rc\rk
                                                      & = & 0.
\eea

$F$ and $G$ are two $U(1)$ fields and $\phi$ and $a$ are the dilaton and
the axion respectively. The later is a pseudoscalar. These can be
combined in a single complex scalar $z=e^{-2\phi}-ia$ ($z$ is $-i$ times
$\lambda$ in reference \cite{kn:STW}) using the self- and anti-self-dual
parts of $F$ and $G$, which makes the duality rotations easier
to describe. In terms of $z$ the action and equations of motion are

\be
S_{SU(4)}=
\int dx^{4}\sqrt{-g}\lk-R+
\superfrac{\partial_{\mu}z\partial^{\mu}\bz}{(z+\bz)^{2}}
-\lc z\lc (F^{+})^{2}+(G^{+})^{2}\rc+ c.c. \rc\rk, \label{eq:action}
\ee

\bea
\nabla_{\mu}(zF^{+\mu\nu}+c.c.)                        & = & 0, \non \\
\nabla_{\mu}(F^{+\mu\nu}-c.c.)                         & = & 0, \non \\
\nabla_{\mu}(zG^{+\mu\nu}+c.c.)                        & = & 0, \non \\
\nabla_{\mu}(G^{+\mu\nu}-c.c.)                         & = & 0, \non \\
\nabla^{2}z-2\frac{(\partial z)^{2}}{(z+\bz)}+
(z+\bz)^{2}\lc (F^{-})^{2}+(G^{-})^{2}\rc              & = & 0, \non \\
\nabla^{2}\bz-2\frac{(\partial \bz)^{2}}{(z+\bz)}+
(z+\bz)^{2}\lc (F^{+})^{2}+(G^{+})^{2}\rc              & = & 0, \non \\
R_{\mu\nu}+\superfrac{\partial_{(\mu}z\partial_{\nu)\bz}}{z+\bz}\: -
                                                       &   &   \non \\
-(z+\bz)\lc F_{\mu\rho}F_{\nu}^{\rho}-\cuarto g_{\mu\nu}F^{2}
+G_{\mu\rho}G_{\nu}^{\rho}-\cuarto g_{\mu\nu}G^{2}\rc  & = & 0.
                                         \label{eq:motion}
\eea

The $SO(4)$ version is obtained by substituting first $G$ by
$\tilde{G}$ defined by

\be
G^{+}_{\mu\nu}=-iz^{-1}\tilde{G}^{+}_{\mu\nu}
\ee

\noindent in the equations of motion. Then it is still necessary to
reverse the sign of the $\tilde{G}$-terms in the action.

It is just a matter of calculation to see that a solution is provided by

\bea
ds^{2}    & = & e^{2U}dt^{2}-e^{-2U}dr^{2}-R^{2}d\Omega^{2},     \non \\
e^{2\phi} & = & e^{2\phi_{0}}\frac{(r+\Sigma)}{(r-\Sigma)},      \non \\
a         & = & a_{0},                                           \non \\
F         & = & Q_{F}\frac{e^{2(\phi-\phi_{0})}}{R^{2}}dt\wedge dr
                -P_{F}\sin{\theta}d\theta\wedge d\phi,           \non \\
G         & = & Q_{G}\frac{e^{2(\phi-\phi_{0})}}{R^{2}}dt\wedge dr
                -P_{G}\sin{\theta}d\theta\wedge d\phi,
\eea

\noindent where the different functions and constant are

\bea
e^{2U}    & = & \superfrac{(r-r_{+})(r-r_{-})}{R^{2}},           \non \\
R^{2}     & = & r^{2}-\Sigma^{2},                                \non \\
\Sigma    & = & e^{-2\phi_{0}}\superfrac{(P_{F}^{2}+P_{G}^{2})
                -(Q_{F}^{2}+Q_{G}^{2})}{2M}
                = \Sigma_{F}+\Sigma_{G},                         \non \\
r_{\pm}   & = & M\pm r_{0},                                      \non \\
r_{0}^{2} & = & M^{2}+\Sigma^{2}-e^{-2\phi_{0}}
                (Q_{F}^{2}+P_{F}^{2}+ Q_{G}^{2}+P_{G}^{2}).
\eea

$Q_{F(G)}$ and $P_{F(G)}$ are the $F(G)$ field electric and magnetic
charges, respectively.
$\Sigma_{F(G)}$ is the $F(G)$ contribution to
the dilaton charge $\Sigma$. For this family of solutions they are not
independent quantities. $\phi_{0}$ and
$a_{0}$ are the asymptotic (constant) value of the dilaton and axion.
Only if the electric and magnetic charges satisfy

\be
Q_{F}P_{F}+Q_{G}P_{G}=0   \label{eq:condition}
\ee

\noindent does the axion equation hold and we have a solution of the
equations of motion.

Motivated by what follows, we define the axion
charge $\Delta$ as a quantity depending on the electric and magnetic
charges as well on the asymptotic value of the dilaton in this way

\be
\Delta=-e^{-2\phi_{0}}\superfrac{(Q_{F}P_{F}+Q_{G}P_{G})}{M}
=\Delta_{F}+\Delta_{G},
\ee

\noindent so we can interpret (\ref{eq:condition}) as the condition of
null axion charge. The charge of the complex scalar is then

\be
\Upsilon=\Sigma-i\Delta= \frac{(P_{F}+iQ_{F})^{2}}{2M}
+\frac{(P_{G}+iQ_{G})^{2}}{2M}=\Upsilon_{F}+\Upsilon_{G},
\ee

\noindent and can correspondingly be expressed in terms on the (complex)
charge of the self-dual parts of $F$ and $G$

\bea
\Gamma_{F(G)}      & = & \medio(Q_{F(G)}+iP_{F(G)}),                 \\
\Upsilon_{F(G)}    & = & -2\frac{(\overline{\Gamma_{F(G)}})^{2}}{M}.
\eea

All these relations are consistent with the definitions of the charges
in terms of the asymptotic behavior ($r\rightarrow \infty$) of the
different fields

\bea
g_{tt}       & \sim & 1-\frac{2M}{r},                           \non \\
\phi         & \sim & \phi_{0}+\frac{\Sigma}{r},\hspace{2cm}
                       (e^{-2\phi}\sim e^{-2\phi_{0}}
                       (1-\frac{2\Sigma}{r})),                  \non \\
a            & \sim & a_{0}-e^{-2\phi_{0}}\frac{2\Delta}{r},    \non \\
z            & \sim & z_{0}-e^{-2\phi_{0}}\frac{2\Upsilon}{r},  \non \\
F_{tr}       & \sim & \frac{Q_{F}}{r},                          \non \\
\star F_{tr} & \sim & i\frac{P_{F}}{r},                         \non \\
F^{+}_{tr}   & \sim & \frac{\Gamma_{F}}{r}.  \label{eq:charges}
\eea

Now
let us briefly describe the main properties of this family of solutions.
Setting $P_{F}=Q_{G}=0$ we get that of reference \cite{kn:KLOPP}
which essentially is one of the solutions in \cite{kn:Gibb} and
\cite{kn:GM} generalized to $\phi_{0}\neq 0$. Reissner-Nordstr\"{o}m
black holes and the charged dilaton black holes described in
\cite{kn:GHS} were already included in \cite{kn:Gibb} and \cite{kn:GM}.
The axion charge was zero in the cases considered there because
the contributions of the fields $F$ and $G$ to it were identically zero:
$\Delta_{F}=\Delta_{G}=0$. Now, in the more general case considered
here both contributions cancel each other: $\Delta_{F}=-\Delta_{G}$.

We remark that our solutions
can not be obtained from the solutions considered in references
\cite{kn:KLOPP}, \cite{kn:Gibb} and \cite{kn:GM} by means of a $SL(2,R)$
transformation of the type we are going to consider in section 3.

Notice that now the metric, dilaton and axion fields
are essentially the same as in those references the only difference
being that we have to replace every single electric or magnetic charge
in their expressions by the orthogonal sum of our pair of charges.
All the properties that depend on the metric (Hawking temperature
etc.) can be found in this way from those of the case $P_{F}=Q_{G}=0$.
This means in particular that we again have spherical black
holes with two horizons at $r=r_{\pm}$ which coincide when $r_{0}=0$.

Black
holes with $r_{0}=0$ are called extreme because they are on the verge of
having a naked singularity. No solution with smaller mass and the same
charges is regular outside a horizon. Since there are no $U(1)$ charged
particles in our theory, this means that the evaporation of any
regular black hole ($r_{0}\geq 0$) should stop when they become extreme.
It was proved in $\cite{kn:KLOPP}$ that the
extreme black holes with $P_{F}=Q_{G}=0$ have unbroken supersymmetries
and saturate the Bogomolnyi-like bound

\be
M^{2}+\Sigma^{2}                                              \geq
e^{-2\phi_{0}}\lp Q_{F}^{2}+P_{G}^{2}\rp.
\ee

In our case, the condition $r_{0}^{2}\geq 0$ gives

\be
M^{2}+\Sigma^{2}+\Delta^{2}                                    \geq
e^{-2\phi_{0}} \lp Q_{F}^{2}+Q_{G}^{2}+P_{F}^{2}+P_{G}^{2}\rp,
\ee

\noindent which is more easily written by using the complex charges
defined above

\be
M^{2}+\mid\Upsilon\mid^{2}                                         \geq
e^{-2\phi_{0}}\lp \mid\Gamma_{F}\mid^{2}+\mid\Gamma_{G}\mid^{2}\rp.
                                  \label{eq:bound}
\ee

We have included the (vanishing) axion charge with the sign with which
it will appear later. This bound is also saturated for extreme black
holes, and so we expect them to have unbroken supersymmetries.
 Our next task will be to prove that this property actually holds
for the more general class of extreme black holes we are dealing with.
We won't derive the Bogomolnyi-like bound (\ref{eq:bound}) from the
supersymmetry algebra here, though.

\section{Supersymmetry of the doubly charged solutions}

Now we want to  consider our solution as the
bosonic part of a solution of the full supersymmetric theory in which
all the fermionic fields are zero. A natural question to ask is whether
this solution is invariant under some local supersymmetry
transformations. The variations of the bosonic fields are proportional
to the fermionic fields and so they are obviously zero. The variations
of the fermionic fields (gravitinos and dilatinos) are proportional to
the bosonic fields and so, only for very special backgrounds there will
be a finite number of transformations leaving them partially invariant,
i. e. obeying $\delta_{\epsilon}\psi_{\mu I}=0$
and $\delta_{\epsilon}\Lambda_{I}=0$ for some of the $SO(4)$ indices
$I$. Finding these transformations is equivalent to finding
the spinors $\eid$ ($\eiu$) that satisfy the differential equations

\bea
\nabla_{\mu}\eid -\frac{i}{4}e^{2\phi}\partial_{\mu}a\eid
-\frac{e^{-\phi}}{2\sqrt{2}}\sigma^{\rho\sigma}\lc
F_{\rho\sigma}\alpha_{IJ}+z^{-1}\tilde{G}_{\rho\sigma}\beta_{IJ}\rc
\gamma_{\mu}\eju                                             & = & 0,\\
-\gamma^{\mu}(\partial_{\mu}\phi+
\frac{i}{2}e^{2\phi}\partial_{\mu}a)\eid
+\frac{e^{-\phi}}{\sqrt{2}}\sigma^{\rho\sigma} \lc
F_{\rho\sigma}\alpha_{IJ}-\bz^{-1}\tilde{G}_{\rho\sigma}\beta_{IJ}\rc
\gamma_{\mu}\eju                                             & = & 0,
\eea

\noindent respectively in the $SO(4)$ formulation that we use here for
convenience.

On the other hand, if some $\eid$s exist that are asymptotically
constant in the limit $r\rightarrow\infty$, we can speak about
asymptotic, rigid (i. e. non-local) supersymmetry and establish
Bogomolnyi-like
bounds concerning the (asymptotically defined) charges of the
background. Our purpose here is to find this kind of solutions among the
family described in section 1. The construction of the (asymptotic)
supersymmetry subalgebra that leaves invariant the states representing
the backgrounds and which is necessary to derive the Bogomolnyi bound
won't be explicitly made here for this would repeat the work done in
\cite{kn:KLOPP}. Only the new basis of supersymmetry will be shown.
Accordingly we
will impose the condition of time independence on the solutions we are
looking for:

\be
\partial_{t}\eid=0.
\ee

We take $a_{0}=0$ for simplicity. After some algebra,
adding and subtracting the equations $\delta_{\epsilon}\psi_{t I}=0$ and
$\delta_{\epsilon}\Lambda_{I}$ we arrive at the equations

\bea
(\partial_{r}e^{U+\phi})\eid                            & = &
-\sqrt{2}e^{-2\phi_{0}}\frac{e^{2\phi}}{R^{2}}\lp
Q_{F}\alpha_{IJ}+iQ_{G}\beta_{IJ} \rp \gamma^{0}\eju,
                                                     \label{eq:uno} \\
(\partial_{r}e^{U-\phi})\eid                            & = &
-i\sqrt{2}\frac{e^{-2\phi}}{R^{2}}\lp
P_{F}\alpha_{IJ}+iP_{G}\beta_{IJ} \rp \gamma^{0}\eju.
                                                     \label{eq:dos}
\eea

Using them in $\delta_{\epsilon}\psi_{r I}=0$,
$\delta_{\epsilon}\psi_{\theta I}=0$
and $\delta_{\epsilon}\psi_{\varphi I}=0$ we obtain

\bea
\partial_{r}(e^{-\medio U}\eid)                              & = & 0,
                                                  \label{eq:tres}    \\
\partial_{\theta}\eid - \frac{i}{2}\partial_{r}(R e^{U})
\gamma^{3}\gamma^{0}\eid                                     & = & 0,
                                                  \label{eq:cuatro}  \\
\partial_{\varphi}\eid - \frac{i}{2}\sin{\theta}\partial_{r}(R e^{U})
\gamma^{2}\gamma^{0}\eid
+\cos{\theta}\gamma^{1}\gamma^{0}\eid                        & = & 0.
                                                  \label{eq:cinco}
\eea

The solution to the first of these equations is

\be
\eid = e^{\medio U}\hat{\eid},
\ee

\noindent where $\hat{\eid}$ is a function of only $\theta$ and
$\varphi$.
The other two equations have the same form in terms of $\hat{\eid}$.
Now, if we apply $\partial_{r}$ to any of them we get the integrability
condition

\be
\partial_{r}^{2}(R e^{U})=0 \Rightarrow R e^{U}=br+d, \label{eq:intcond}
\ee

\noindent which is tantamount to saying that it has to be possible to
write the metric in isotropic form

\be
ds^{2}=e^{2U}dt^{2}-e^{-2U}[d\rho^{2}+\rho^{2}d\Omega^{2}].
\ee

This happens only when $r_{0}=0$, the extreme case. From now on we
sill write

\be
\partial_{r}(R e^{U})=1.
\ee

Now it is easy to check that the two last angular
equations are solved by

\be
\hat{\eid} = e^{\frac{i}{2}\gamma^{3}\gamma^{0}\theta}
e^{\frac{i}{2}\gamma^{1}\gamma^{0}\varphi}\eio,
\ee

\noindent where $\eio$ is a constant spinor. The exponentials can be
calculated using the explicit expressions for gamma matrices written in
appendix A. The result is

\be
e^{\frac{i}{2}\gamma^{i}\gamma^{0}x}=
\cos{\frac{x}{2}}+i\gamma^{i}\gamma^{0}\sin{\frac{x}{2}}.
\ee

Then, if a solution exists, it must have the form

\be
\eid(r,\theta,\varphi)=
e^{\medio U}e^{\frac{i}{2}\gamma^{3}\gamma^{0}\theta}
e^{\frac{i}{2}\gamma^{1}\gamma^{0}\varphi}\eio. \label{eq:solution}
\ee

Observe that, as we wanted, these spinors are asymptotically constant
\footnote{The angular dependence can be eliminated by a change of
coordinates. Compare, for instance, our solution with equation $(105)$
of reference \cite{kn:KLOPP}}.

So far, we have solved the equations involving only the negative
chirality
spinors. Now we have to solve the equations relating positive and
negative chirality spinors, namely (\ref{eq:uno}) and (\ref{eq:dos}) and
also the Majorana condition which, with our conventions reads

\be
\chi_{I}=\gamma^{2}(\chi^{I})^{\ast}.         \label{eq:Majorana}
\ee

Making the same choice of $\alpha_{IJ}$ and $\beta_{IJ}$ as in
reference \cite{kn:KLOPP}, consistency between equations
(\ref{eq:uno}), (\ref{eq:dos}) and (\ref{eq:Majorana}) is achieved if

\bea
\mid \sqrt{2}\frac{e^{2(\phi-\phi_{0})}}{R^{2}}(Q_{F}+iQ_{G}) \mid
& = &
\partial_{r}e^{U+\phi},                                              \\
\mid \sqrt{2}\frac{e^{-2\phi}}{R^{2}}(P_{F}+iP_{G}) \mid
& = &
\partial_{r}e^{U-\phi},
\eea

\noindent which is true only in the extreme case. Within this family of
solutions, only extreme black holes can have unbroken supersymmetries.
Now let's define for ($U\neq\pm\phi$) the two complex phases

\bea
\xi  & = & \superfrac{Q_{F}+iQ_{G}}{(Q_{F}^{2}+Q_{G}^{2})^{\medio}}, \\
\eta & = & \superfrac{P_{F}+iP_{G}}{(P_{F}^{2}+P_{G}^{2})^{\medio}},
\eea

\noindent where we choose the positive branch of the square root. These
complex phases were just plus or minus signs in the cases covered in
reference \cite{kn:KLOPP}. In the $I,J=1,2$ sector we have

\bea
\partial_{r}e^{U+\phi}\lp \epsilon_{1}+
\xi^{\ast}\gamma^{0}\epsilon^{2} \rp                 & = &  0,   \non \\
\partial_{r}e^{U-\phi}\lp \epsilon_{1}+
i\eta^{\ast}\gamma^{0}\epsilon^{2} \rp               & = &  0,
\eea

\noindent and in the $I,J=3,4$ sector we have

\bea
\partial_{r}e^{U+\phi}\lp \epsilon_{3}+
\xi\gamma^{0}\epsilon^{4} \rp                 & = &  0,   \non \\
\partial_{r}e^{U-\phi}\lp \epsilon_{3}+
i\eta\gamma^{0}\epsilon^{4} \rp               & = &  0.
\eea

Notice that these equations have solutions only for the {\it pairs} of
spinors $\epsilon_{1},\epsilon_{2}$ and$/$or
$\epsilon_{3},\epsilon_{4}$. The four spinors have to be of the form
given by (\ref{eq:solution}), the difference being the constant
spinor we pick. The previous equations are, then, two relations between
$\epsilon_{1_{0}}$ and $\epsilon_{2_{0}}$ and two relations between
$\epsilon_{3_{0}}$ and $\epsilon_{4_{0}}$. Both relations must be
compatible.

Then, unbroken supersymmetry in the $1,2$ sector means

\bea
\epsilon_{1} & = & -\xi^{\ast}\gamma^{0}\epsilon^{2},            \non \\
i\eta^{\ast} & = &  \xi^{\ast},
\eea

\noindent which implies, for some positive finite constant $\mu$

\bea
Q_{F} & = &  \mu P_{G},                                          \non \\
Q_{G} & = & -\mu P_{F},
\eea

\noindent that is, $\Delta=0$. Only $\epsilon_{1_{0}}$ or
$\epsilon_{2_{0}}$ can be
chosen arbtrarily (but subject to the chirality constraint). This means
two complex (four real) arbitrary constants, that is, unbroken $N=1$
supersymmetry.

In the $3,4$ sector, unbroken supersymmetry means

\bea
\epsilon_{3} & = & -\xi\gamma^{0}\epsilon^{4},                  \non \\
i\eta & = &  \xi,
\eea

\noindent which implies, for some positive finite constant $\mu$

\bea
Q_{F} & = & -\mu P_{G},                                        \non \\
Q_{G} & = &  \mu P_{F}.
\eea

We
obtain again the condition $\Delta=0$ and unbroken $N=1$ supersymmetry.

Note that for $U\neq\pm\phi$ only one of the pairs of
supersymmetries $1,2$ or $3,4$ can be unbroken at once. We can only have
unbroken  $N=1$ supersymmetry.

We can define a new basis for the supersymmetries

\bea
\epsilon^{\;\;ij}_{\xi}                                           & = &
\epsilon_{j}+\xi\gamma^{0}\epsilon^{i},                         \non \\
\epsilon_{\xi ij}                                                 & = &
\epsilon^{j}+\xi\gamma^{0}\epsilon_{i},                         \non \\
\epsilon^{\;\;ij}_{\xi^{\ast}}                                    & = &
\epsilon_{j}+\xi^{\ast}\gamma^{0}\epsilon^{i},                  \non \\
\epsilon_{\xi^{\ast}ij}                                           & = &
\epsilon^{j}+\xi^{\ast}\gamma^{0}\epsilon_{i},
\eea

\noindent where the pair $ij$ takes the values $12$ and $34$.
When $sgn(Q_{F})=sgn(P_{G})$ (which implies $sgn(Q_{G})=-sgn(P_{F})$ for
$\Delta$ to be zero) $\epsilon^{\;\;12}_{\xi}$ and $\epsilon_{\xi12}$
are unbroken. If $sgn(Q_{F})=-sgn(P_{G})$ ($sgn(Q_{G})=sgn(P_{F})$)
then $\epsilon^{\;\;34}_{\xi^{\ast}}$ and $\epsilon_{\xi^{\ast}12}$ are
unbroken. Obviously, the basis is different for each specific case.

If $U=+\phi$, then $P_{F}=P_{G}=0$ (purely electric case). We have only
one relation between spinors in each sector which implies no
extra compatibility condition on the charges. This means that we have
unbroken supersymmetry both in the $1,2$ and in the $3,4$ sectors, and,
in consequence unbroken $N=2$ supersymmetry.
The $\epsilon^{\;\;ij}_{\xi}$ and
$\epsilon_{\xi ij}$ $ij=12,34$ are the unbroken supersymmetries in this
case.

Let us summarize the results obtained in this section. Doubly charged
extreme solutions with $U\neq\pm\phi$ are $N=1$ supersymmetric.
The cases $U=\pm\phi$ (purely electric and purely magnetic
respectively) are $N=2$ supersymmetric. No non-extreme
solution is supersymmetric.

\section{SL(2,R) rotation of doubly charged dilaton black holes}

In reference \cite{kn:STW} it was found that the equations of motion of
the action (\ref{eq:action})\footnote{The inclusion of the second vector
field $G$ doesn't make any qualitative difference.} were
almost invariant under an $SL(2,R)$ group of transformations generated
by performing alternatively Peccei-Quinn shifts of the axion by a
constant $a\rightarrow a+c$

\be
z \rightarrow z-i\beta,
\ee

\noindent and the duality transformation
which in absence of axion is the transformation $\phi\rightarrow
-\phi$ that trades electric for magnetic solutions

\bea
z     & \rightarrow & 1/z,                                      \non \\
F^{+} & \rightarrow & -izF^{+},                                 \non \\
F^{-} & \rightarrow & i\bz F^{-}.
\eea

The action of a general $SL(2,R)$ transformation on $z$ and
$F$ is\footnote{There is an ambiguity in the sign of $(i\gamma+\delta)$
corresponding to the fact that if $M\in SL(2,R)$, $M$ and $-M$ have the
same action on $z$. Whatever sign we choose we get the same result.}

\bea
z     & \rightarrow & \superfrac{\alpha z-i\beta}{i\gamma z+\delta},
                      \,\,\,\, \alpha\delta-\beta\gamma=1,       \non \\
F^{+} & \rightarrow & -(i\gamma z+\delta)F^{+}. \label{eq:duality}
\eea

As it is explained in reference \cite{kn:Sendual} this is an exact
symmetry of the equations of motion because the offending extra term
that appears in the last equation of motion (\ref{eq:motion}) is
proportional to

\be
F_{(\alpha|\rho|}\star F_{\beta)}^{\;\rho}-
\cuarto g_{\alpha\beta}F\star F,
\ee

\noindent (and an
analogous term for $G$) which vanishes identically in four dimensions.
(One has to calculate each component of this expression or use the
Newmann-Penrose formalism to prove it.) The existence of this group
of exact invariances was already known in the context of $N=4$, $d=4$
ungauged supergravity (the $SU(1,1)$ group of reference \cite{kn:CSF}).
In this respect, the novelty in \cite{kn:Sendual} is the inclusion of
more terms in the action coming from the compactification of the 6
extra dimensions of heterotic superstring theory on a torus.

In this section we are going to use the reduced subset of $SL(2,R)$
transformations that was used in reference \cite{kn:STW} to generate
backgrounds with non-trivial axion field. They consist of a shift of the
axion by a constant $\beta=c$ followed by the duality transformation
(\ref{eq:duality}) and a rescaling by the normalization constant $N$

\bea
z     & \rightarrow & z^{\prime}=\superfrac{-i
                      (-N)^{\medio}}{iN^{-\medio}z+cN^{-\medio}},
                                                              \non \\
F^{+} & \rightarrow & F^{+\prime}=-(iN^{-\medio}z+cN^{-\medio})F^{+},
                                                              \non \\
F^{-} & \rightarrow & F^{-\prime}=-(-iN^{-\medio}bz+cN^{-\medio})F^{-},
\eea

\noindent and the same for $G$.
 Using our definitions of the charges
(\ref{eq:charges}) we see that the transformations described above act
on the charges and asymptotic values of any background in the following
way

\newcommand{\eso}{(a_{0}+c)^{2}e^{2\phi_{0}}+e^{-2\phi_{0}}}
\newcommand{\lootro}{(a_{0}+c)^{2}e^{2\phi_{0}}-e^{-2\phi_{0}}}

\bea
a_{0}^{\prime}          & = & \superfrac{-N e^{2\phi_{0}}}{\eso}
                              (a_{0}+c),
                                                                 \non \\
e^{-2\phi_{0}^{\prime}} & = & \superfrac{N}{\eso},               \non \\
Q_{F}^{\prime}          & = & \frac{-(a_{0}+c)}{N^{\medio}}Q_{F}
                              +\frac{e^{-2\phi_{0}}}{N^{\medio}}
                                P_{F},                           \non \\
P_{F}^{\prime}          & = & -\frac{e^{-2\phi_{0}}}{N^{\medio}}
                                Q_{F}
                              -\frac{-(a_{0}+c)}{N^{\medio}}P_{F},
                                                                 \non \\
\Sigma_{F}^{\prime}     & = & \superfrac{\lootro}{\eso}\Sigma_{F}
                             -\superfrac{2(a_{0}+c)}{\eso}
                             \Delta_{F},
                                                                 \non \\
\Delta_{F}^{\prime}     & = & \superfrac{2(a_{0}+c)}{\eso}
                              \Sigma_{F}
                              +\superfrac{\lootro}{\eso}\Delta_{F}.
                                                         \label{eq:rot}
\eea

Similar expressions hold for $G$-charges.

In \cite{kn:STW} $N$ was chosen to preserve
$\mid\Gamma_{F}\mid^{2}$. However, what we want to preserve is actually
$e^{-2\phi_{0}}\mid\Gamma_{F}\mid^{2}$ because
that is what appears in the bound (\ref{eq:bound}).
For any value of $N$, $\mid\Upsilon_{F(G)}\mid^{2}$ and the combination
$e^{-2\phi_{0}}\mid\Gamma_{F(G)}\mid^{2}$ are invariant under
(\ref{eq:rot}). In fact, these transformations can be written as
rotations

\bea
e^{-\phi_{0}^{\prime}}\Gamma^{\prime}                              & = &
e^{-i\alpha}(e^{-\phi_{0}}\Gamma),                               \non \\
\Upsilon^{\prime}                                                  & = &
e^{+2i\alpha}\Upsilon,                                           \non \\
\sin{\alpha}                                                       & = &
\superfrac{e^{-\phi_{0}}}{(\eso)^{\medio}},                      \non \\
\cos{\alpha}                                                       & = &
\superfrac{-(a_{0}+c)}{(\eso)^\medio{}}\;\;.
\eea

Therefore, the Bogomolnyi bound (\ref{eq:bound}) is invariant.
If we rotate a background in which
it is saturated, we will obtain another one with the same property. The
supersymmetry property associated to it should also be preserved.

In fact, it doesn't take much effort to see that the Bogomolnyi
bound is invariant under the full $SL(2,R)$ group. A a consequence, it
should
transform supersymmetric backgrounds into supersymmetric backgrounds.
Here we will focus on the subset of transformations described above
and will prove the supersymmetry of the transformed solutions in the
next section. The general case is not much more illuminating and should
follow straightforwardly. The results will be presented
elsewhere.

The Bogomolnyi bound can also be interpreted as a condition of
equilibrium of forces. The electric-magnetic dual rotations
preserve this equilibrium not only by permutating the charges. Notice
the curious interplay between dilaton and
electromagnetic forces. The scaling of the electric and magnetic charges
in a $SL(2,R)$ transformation is absorbed in the scaling of
$e^{\phi_{0}}$, the string coupling constant at infinity.
The existence of doubly charged multi-extreme-black-hole solutions in
which the Bogomolnyi bound is clearly seen as a condition of
equilibrium of forces is very likely. We won't try to study them here.

\section{Supersymmetry of the rotated solutions}

Here we will present the calculations quite schematically for we
followed exactly the same steps as in section 2 to find the $N=4$
Killing spinors. The
equations analogous to (\ref{eq:uno}), (\ref{eq:dos}), (\ref{eq:tres}),
(\ref{eq:cuatro}) and (\ref{eq:cinco}) are respectively

\bea
(\partial_{r}e^{U+\phi})\eid                                 & = &
\sqrt{2}e^{i\alpha}\frac{e^{2(\phi-\phi_{0})}}{R^{2}}\lp
Q_{F}\alpha_{IJ}+iQ_{G}\beta_{IJ} \rp \gamma^{0}\eju,
                                                  \label{eq:unop}    \\
(\partial_{r}e^{U-\phi})\eid                                 & = &
i\sqrt{2}e^{i\alpha}\frac{e^{-2\phi}}{R^{2}}\lp
P_{F}\alpha_{IJ}+iP_{G}\beta_{IJ} \rp \gamma^{0}\eju,
                                                  \label{eq:dosp}    \\
\partial_{r}(e^{-\medio (U+i\alpha)}\eid)                    & = & 0,
                                                  \label{eq:tresp}   \\
\partial_{\theta}\eid - \frac{i}{2}\partial_{r}(R e^{U})
\gamma^{3}\gamma^{0}\eid                                     & = & 0,
                                                  \label{eq:cuatrop} \\
\partial_{\varphi}\eid-                                      &   &
                                                                \non \\
-\frac{i}{2}\sin{\theta}\partial_{r}(R e^{U}) \gamma^{2}\gamma^{0}\eid+
                                                             &   &
                                                                \non \\
+\cos{\theta}\gamma^{1}\gamma^{0}\eid                        & = & 0,
                                                  \label{eq:cincop}
\eea

\noindent where we have expressed all the primed (transformed) fields
and charges
in terms of the unprimed (original) ones and $\alpha$ is the argument of
$c+ie^{-2\phi}$. We have almost recovered the equations of section 2\ !
The only evidence of the existence of an axion is the complex
function $e^{i\alpha}$.

Equation (\ref{eq:tresp}) is solved by

\be
\eid = e^{\medio(U+i\alpha)}\hat{\eid},
\ee

\noindent so the supercovariant spinors are different from those of the
original solutions, the difference being the complex ($r$-dependent)
phase $e^{\frac{i}{2}\alpha}$. Further comments on the presence of this
phase will be made in the last section.

The integrability condition of equations (\ref{eq:cuatrop}),
(\ref{eq:cincop}) and (\ref{eq:tresp}) is again the extremality
condition (\ref{eq:intcond}). The solution to
(\ref{eq:tresp}), (\ref{eq:cuatrop}) and (\ref{eq:cincop}) is

\be
\eid =
e^{\medio (U+i\alpha)}
e^{\frac{i}{2}\gamma^{3}\gamma^{0}\theta}
e^{\frac{i}{2}\gamma^{1}\gamma^{0}\varphi}
\epsilon_{I_{0}}.
\ee

Considering now equations (\ref{eq:unop}) and (\ref{eq:dosp}),
we split
them into the $1,2$ and $3,4$ sectors etc. (in this case we choose the
negative branch of the square root) we arrive at

\bea
\partial_{r}e^{U+\phi}\lp \epsilon_{1}+
e^{i\alpha}\xi^{\ast}\gamma^{0}\epsilon^{2} \rp      & = &  0,  \non \\
\partial_{r}e^{U-\phi}\lp \epsilon_{1}+
i e^{i\alpha}\eta^{\ast}\gamma^{0}\epsilon^{2} \rp   & = &  0,
\eea

\noindent for the $I,J=1,2$ sector, and for the $I,J=3,4$ sector we have

\bea
\partial_{r}e^{U+\phi}\lp \epsilon_{3}+
e^{i\alpha}\xi\gamma^{0}\epsilon^{4} \rp             & = &  0,  \non \\
\partial_{r}e^{U-\phi}\lp \epsilon_{3}+
i e^{i\alpha}\eta\gamma^{0}\epsilon^{4} \rp          & = &  0.
\eea

The discussion of broken and unbroken supersymmetries is the same we
made in section 2, including the fact that unbroken supersymmetry
implies $\Delta=0$ (not $\Delta^{\prime}=0$!). We won't repeat it here.
However, notice that whether a particular supersymmetry is broken or
unbroken does not depend on the signs of the {\it actual} (primed)
charges of the solutions but on the {\it original}
(unprimed) charges, which are linear combinations of the actual ones.

\section{Conclusions}

In this paper we have described a new class of solutions to the string
effective action with vanishing axion field and found that the extreme
ones have
unbroken supersymmetries. This provides a more general example in which
supersymmetry plays the role of cosmic censor in static asymptotically
flat spaces. The related Bogomolnyi-like bound has been displayed,
including the axion charge.

Furthermore, we have studied the effect of the $SL(2,R)$ group of
electric-magnetic duality rotations that leave the equations of motion
invariant on the supersymmetry properties and we have seen that they
leave invariant the form of the Bogomolnyi bound (with the axion charge
included), giving us a hint that
they preserve the unbroken supersymmetries. We have performed explicitly
the rotation of this class of solutions under a subset of the whole
$SL(2,R)$ generating a non-trivial axion field and learned that the
transformed solutions have the same number of unbroken supersymmetries
the original solutions had. We conjecture that this will be true for any
solution and any $SL(2,R)$ duality rotation.

The presence of the factor $e^{i\alpha}$ in the supercovariantly
constant spinors, which is the only trace of the presence of an axion
is a remarkable fact. In reference \cite{kn:Tod}, all the backgrounds
admitting $N=2$ supercovariantly constant spinors were found. Apart from
the plain waves, they were nothing but generalizations of the
Israel-Wilson-Perjes \cite{kn:IWP} class of matrics including
charged, rotating dust. They can be described in terms of a single
complex
function $V$, from which the (unique) $U(1)$ field is also obtained.
$V$ was in the context of \cite{kn:Tod} the product of the pair of
supercovariantly constant $SL(2,C)$ two component spinors, and it would
be $e^{U+i\alpha}$ in our case.
Some of these metrics admit also $N=4$ supercovarianty constant spinors.
Of course, the vector fields are different, and we have to add dilaton
and axion. For instance, the
metrics of the whole class of extreme electric and magnetic dilaton
black
holes described in \cite{kn:GHS} for any value of the parameter $a$ are
described in \cite{kn:Tod}. The vector field has to be rescaled by a
power of the function $V$, which is purely real or imaginary in this
case, and the dilaton has to be identified with the appropriate
function of $V$ ($e^{U}$). We have

\bea
V       & = & e^{\frac{\phi}{a}},                                    \\
F^{Tod} & = & (\medio(1+a^{2}))^{-\medio}e^{-a\phi}F^{GHS},
\eea

\noindent where $V$ has to satisfy the equation

\be
\nabla^{2}V^{-(1+a^{2})}=0,
\ee

\noindent so $V^{-(1+a^{2})}$ is a harmonic function\footnote{This
permits us to find new multi-black-hole solutions}. However, basically
only for $a=1$ \cite{kn:KLOPP} we can embed these solutions in a theory
with local supersymmetry\footnote{R. Kallosh and A. Van Proeyen, private
communication.}

In this case $V$ is essentially
the dilaton, even though there is no dilaton in $N=2$ supergravity.
In the case at hand we have a complex $V$, the imaginary part
reflecting the fact that we have an axion. There is no axion in $N=2$,
either, and solutions with complex $V$ in \cite{kn:Tod} have rotation in
general and have not diagonal metrics. However, there might be a way of
trading rotation by axion, so we can stablish a (formal) connection
between two kinds of systems otherwise very different.

We haven't explored to its full extent the implications that the
preservation of unbroken supersymmetry by $SL(2,R)$ transformations can
have if, as conjectured in \cite{kn:Sendual}, these are symmetries of
the exact string theory and the spectrum of charged black holes is
related to the spectrum of excitations of fundamental strings. However,
our work seems to support this conjecture by telling us that the
supersymmetric structure of the spectrum (supermultiplets) would also be
preserved by these transformations. Further work is needed to
give a final answer to these problems.

\appendix

\section{Coventions}

Here we specify some of the conventions in \cite{kn:KLOPP}. Our choice
of gamma matrices is

\bea
\gamma^{0} & = & \lp \ba{cc}
                            0 & i \\
                            i & 0 \\
                                          \ea \rp ,
\hspace{2cm}
\gamma^{i}  =  \lp \ba{cc}
                            0           & -i\sigma^{i} \\
                            i\sigma^{i} & 0            \\
                                                        \ea \rp \non \\
\gamma_{5} & = & \lp \ba{cc}
                           -1 & 0 \\
                            0 & 1 \\
                                          \ea \rp ,
\hspace{2cm}
{\cal C}  =  \lp \ba{cc}
                           i\sigma^{2}  & 0            \\
                           0            & -i\sigma^{2} \\
                                                        \ea \rp
=i\gamma^{2}\gamma^{0}.
\eea

Our choice of vierbeins is

\be
e_{t}^{0}=e^{U},\,\, e_{r}^{1}=e^{-U},\,\, e_{\theta}^{2}=R,\,\,
e_{\varphi}^{3}=R\sin{\theta},
\ee

so the covariant derivative acting on spinors has the following
components

\bea
\nabla_{t}\epsilon       & = & \cuarto(\partial_{r}e^{2U})
                               \gamma^{1}\gamma^{0} \epsilon,   \non \\
\nabla_{r}\epsilon       & = & \partial_{r} \epsilon,           \non \\
\nabla_{\theta}\epsilon  & = & \partial_{\theta} \epsilon
                               -\medio e^{U} \partial_{r}R
                               \gamma^{1}\gamma^{2}\epsilon,    \non \\
\nabla_{\varphi}\epsilon & = & \partial_{\varphi} \epsilon
                              -\medio e^{U} \partial_{r}R\sin{\theta}
                               \gamma^{1}\gamma^{3}\epsilon
                              -\medio e^{U} \cos{\theta}
                               \gamma^{2}\gamma^{3}\epsilon.
\eea

Obviously, $t,r,\theta$ and $\varphi$ are curved indices, and $0,1,2$
and $3$ are flat.

\section*{Acknowledgements}

The possibility of the existence of doubly charged solutions
was suggested to the author by A. Linde in a private conversation.
He is also indebted to R. Kallosh and A. Van Proeyen for many helpful
discussions from which I greatly benefited. Finally, the author is
the most grateful to M. M. Fern\'andez for her support and encouragement.

This work has
been partially supported by a Spanish Government MEC postdoctoral grant.


\end{document}